\documentclass[11pt]{article}
\usepackage{amsmath,amssymb,cite}

\makeatletter
\@addtoreset{equation}{section}
\makeatother

\def\ben{\begin{equation}}
\def\een{\end{equation}}

\def\nn{\nonumber} \def\bd{\begin{document}} \def\ed{\end{document}}
\def\ds{\documentstyle} \let\fr=\frac \let\bl=\bigl \let\br=\bigr
\let\Br=\Bigr \let\Bl=\Bigl
\let\bm=\bibitem
\let\na=\nabla
\let\pa=\partial \let\ov=\overline
\newcommand{\be}{\begin{equation}}
\newcommand{\ee}{\end{equation}}
\def\ba{\begin{array}}
\def\ea{\end{array}}
\def\ft#1#2{{\textstyle{\frac{\scriptstyle #1}{\scriptstyle #2} } }}
\def\fft#1#2{{\frac{#1}{#2}}}
\def\del{\partial}
\def\vp{\varphi}
\def\sst#1{{\scriptscriptstyle #1}}
\def\oneone{\rlap 1\mkern4mu{\rm l}}
\def\td{\tilde}
\def\wtd{\widetilde}
\def\ie{{\it i.e.\ }}
\def\dalemb#1#2{{\vbox{\hrule height .#2pt
        \hbox{\vrule width.#2pt height#1pt \kern#1pt
                \vrule width.#2pt}
        \hrule height.#2pt}}}
\def\square{\mathord{\dalemb{6.8}{7}\hbox{\hskip1pt}}}
\newcommand{\ho}[1]{$\, ^{#1}$}
\newcommand{\hoch}[1]{$\, ^{#1}$}
\newcommand{\bea}{\begin{eqnarray}}
\newcommand{\eea}{\end{eqnarray}}
\newcommand{\ra}{\rightarrow}
\newcommand{\lra}{\longrightarrow}
\newcommand{\Lra}{\Leftrightarrow}
\newcommand{\bp}{\tilde \beta^\prime}
\newcommand{\tr}{{\rm tr} }
\newcommand{\Tr}{{\rm Tr} }
\def\0{{\sst{(0)}}}
\def\1{{\sst{(1)}}}
\def\2{{\sst{(2)}}}
\def\3{{\sst{(3)}}}
\def\4{{\sst{(4)}}}
\def\5{{\sst{(5)}}}
\def\6{{\sst{(6)}}}
\def\7{{\sst{(7)}}}
\def\8{{\sst{(8)}}}
\def\n{{\sst{(n)}}}
\def\cA{{{\cal A}}}
\def\cB{{{\cal B}}}
\def\cF{{{\cal F}}}
\def\cG{{{\cal G}}}
\def\cH{{{\cal H}}}
\def\cL{{{\cal L}}}

\def\tV{\widetilde V}
\def\tW{\widetilde W}
\def\tH{\widetilde H}
\def\tE{\widetilde E}
\def\tF{\widetilde F}
\def\tA{\widetilde A}
\def\im{{{\rm i}}}
\def\tY{{{\wtd Y}}}
\def\ep{{\epsilon}}
\def\vep{{\varepsilon}}
\def\bD{{{\bar D}}}
\def\R{{{\mathbb R}}}
\def\C{{{\mathbb C}}}
\def\H{{{\mathbb H}}}
\def\CP{{{\mathbb C}{\mathbb P}}}
\def\RP{{{\mathbb R}{\mathbb P}}}
\def\Z{{{\mathbb Z}}}
\def\bA{{{\mathbb A}}}
\def\bB{{{\mathbb B}}}
\def\bC{{{\mathbb C}}}
\def\bD{{{\mathbb D}}}
\def\bE{{{\mathbb E}}}
\def\bZ{{{\mathbb Z}}}
\def\Re{{{\frak{Re}}}}
\def\Im{{{\frak{Im}}}}
\def\cosec{{\,\hbox{cosec}\,}}
\def\Gm{{\Gamma_{\!\! -}}}
\def\Gp{{\Gamma_{\!\! +}}}
\def\stan{{standard }}
\def\nonstan{{supernumerary }}
\def\p{{\partial}}
\def\kdel#1{{\fft{\del}{\del#1}}}

\def\bog{{Bogomolny }}
\def\om{{\omega}}

\newcommand\w[1]{\\[0.#1cm]}
\def\eq#1{(\ref{#1})}
\def\c{{\gamma}}

\newcommand{\damtp}{\it DAMTP, Centre for Mathematical Sciences,
 Cambridge University,\\  Wilberforce Road, Cambridge CB3 OWA, UK}

\newcommand{\tamphys}{\it George and Cynthia Woods Mitchell  Institute
for Fundamental Physics and Astronomy,\\
Texas A\&M University, College Station, TX 77843, USA}

\newcommand{\auth}{
G.W. Gibbons{$^1$}, C.N. Pope{$^{1,2}$} and E. Sezgin{$^2$}
}

\begin{document}

\begin{flushright}
\hfill{
MIFP-08-17}\\
\end{flushright}

\begin{center}

{\large {\bf The General Supersymmetric Solution of Topologically Massive
   Supergravity}}

\vspace{25pt}

\auth

\vspace{10pt}
{$^1$}{\damtp}

\vspace{10pt}
{$^2$}{\tamphys}

\vspace{25pt}

\underline{ABSTRACT}

\end{center}

  We find the general fully non-linear
solution of topologically massive supergravity
admitting a Killing spinor.  It is of plane-wave type, with a null
Killing vector field.  Conversely, we show that all solutions with a
null Killing vector are supersymmetric for one or the other choice of
sign for the Chern-Simons coupling constant $\mu$. If $\mu$ does not take
the critical value $\mu=\pm 1$, these solutions are asymptotically regular
on a Poincar\'e patch, but do not admit a smooth global compactification with
boundary $S^1\times\R$.  In the critical case, the solutions have a
logarithmic singularity on the boundary of the Poincar\'e patch.  We derive
a Nester-Witten identity, which allows us to identify the associated
charges, but we conclude that the presence of the
Chern-Simons term prevents us from making a statement about their positivity.
The Nester-Witten procedure is applied to the BTZ black hole.

\vspace{15pt}

\thispagestyle{empty}

\pagebreak
\setcounter{page}{1}

\tableofcontents

\addtocontents{toc}{\protect\setcounter{tocdepth}{2}}

\section{Introduction}

 There has been considerable interest recently
\cite{comgrav1,comgrav2,comgrav3,comgrav4,comgrav5,comgrav6}
in topologically massive
gravity with a cosmological constant in $2+1$ spacetime dimensions.  In
this paper, we study the supergravity version \cite{DeserKay,Deser}.  In
particular, we find all solutions admitting a Killing spinor, and we
discuss the Nester-Witten identity, and the extent to which it allows
one to prove a positivity result for the total energy.  It should
be noted that there are different definitions for the energy and the
angular momentum for an asymptotically AdS$_3$ solution of the
topologically massive theory.  In this paper, we shall take the view
that energy should be measured above that of a fiducial background
metric.  This is consistent with supersymmetry, but differs from various
holographic constructions \cite{0506176,0508218}.
We find that while the standard Nester-Witten
procedure does allow us to identify asymptotic charges,
the higher-derivative
character of the Chern-Simons term prevents one from making a general
positivity statement.

   The organisation of the paper is as follows.  In section 2, we review
topologically massive supergravity, and its supersymmetry transformations.
In section 3, we find the general solution admitting a Killing spinor, and
show that for suitable choice of sign of the Chern-Simons coupling
constant $\mu$, every solution with a null Killing vector admits a Killing
spinor. Solutions with a null Killing vector exist for all non-zero  values
of $\mu$, and so far as {\sl local}  degrees of freedom are concerned,
we conclude that their number is the same for all non-vanishing $\mu$.
By ``local degrees of freedom,'' we mean that the number of solutions
in an open neighbourhood of any point in the manifold is the same for
all finite values of $\mu$.  However, the imposition of boundary
conditions at infinity, so as, for instance, to fit them into representations
of the isometry group $SO(2,2)$, does depend on the value of $\mu$, since
in the critical case $\mu=\pm1$, there are logarithmic terms at infinity.
For detailed discussions of this issue, see
\cite{comgrav1,comgrav2,comgrav3,comgrav4,comgrav5,comgrav6}.

   We also
show that all solutions with a null Killing vector are of Kerr-Schild form
and that they are {\it universal} in the sense of \cite{cogihepo}; that is, no
symmetric conserved tensors can be constructed  from the metric and its
derivatives, other than the metric and the Cotton tensor.
This shows that  these metrics are unchanged in form by
quantum corrections to arbitrary order in perturbation theory.

  In section 4, we discuss global issues and the
the asymptotic structure of the solutions.
Those for generic values of $\mu$ are asymptotically regular in a
Poincar\'e patch, but in the critical cases $\mu=\pm 1$ there are
logarithmic singularities.  In no case do the solutions admit a smooth
conformal compactification with conformal boundary $S^1\times \R$.  In
section 5, we give a Nester-Witten formula for the total energy, and
find that the bulk term contains a contribution from the Cotton tensor.
This precludes a general conclusion about the sign of the energy.  Finally,
we exhibit the Killing spinors of the BTZ
vacuum ($M=J=0$) and maximally rotating ($J+M=0$) BTZ black holes.  Further,
we show that there is a single Nester-Witten charge, which is equal to
$M +J$.

\section{$N=1$ Topologically Massive Supergravity}

Simple topological massive supergravity, which is the sum of simple
supergravity and a gravitational super Chern-Simons action, was
constructed by Deser and Kay \cite{DeserKay}.  It was generalised
to include a cosmological term by Deser \cite{Deser}. The total
Lagrangian, in our conventions, is given by
\bea
e^{-1} \cL &=& R + 2 m^2 - 2 {\bar\psi}_\mu
\c^{\mu\nu\rho}D_\nu\psi_\rho
 - m {\bar\psi}_\mu \c^{\mu\nu}\psi_\nu \nn\w2
&& -\ft14
   \mu^{-1}\,\varepsilon^{\mu\nu\rho}\left( R_{\mu\nu}{}^{ab} \omega_{\rho
   ab}+ \ft23 \omega_\mu^{ab} \omega_{\nu b}{}^c\omega_{\rho ca} \right)
   -\mu^{-1} {\bar R}^\mu \c_\nu\c_\mu R^\nu\,,\label{1}
\eea
where we
have set the gravitational coupling constant equal to one, and used the
following curvatures
\bea
R_{\mu\nu}{}^{ab} &=& \partial_\mu
\omega_\nu^{ab}+ \omega_\mu^{ac} \omega_{\nu c}{}^b- (\mu
\leftrightarrow \nu)\,,\w2
R^\mu &=& \varepsilon^{\mu\nu\rho}
 D_\nu(\omega) \psi_\rho\,.
\eea

It is important to note that the spin connection is {\it not} an
independent field, but rather it is given by
\be
\omega_{\mu ab} = \omega_{\mu ab}(e) +
\ft12 ({\bar\psi}_\mu\c_a\psi_b -{\bar\psi}_\mu\c_b\psi_a +
{\bar\psi}_a\c_\mu\psi_b)\,,
\ee
where $\omega_{\mu ab}(e)$ is the standard spin connection that solves the
equation $D_\mu(\omega(e))e_\nu^a = \partial_\mu e_\nu^a
-\Gamma_{\mu\nu}^\rho(g) + \omega_\mu^{ab}(e) e_{\nu b}=0$, with
$\Gamma_{\mu\nu}^\rho(g)$ representing the standard Christoffel
connection. The dreibein also satisfies the metricity condition $D_\mu
e_\nu{}^b = \partial_\mu e_\nu^a -\Gamma_{\mu\nu}^\rho e_\rho^a +
\omega_\mu^{ab}e_{\nu b}=0$, which also serves as the definition of
$\Gamma_{\mu\nu}^\rho$, which, unlike $\Gamma_{\mu\nu}^\rho(g)$, has
torsion.

  The action are is
invariant under the local supersymmetric transformations
\bea
\delta e_\mu^a &=& {\bar\epsilon}\c^a\psi_\mu\,,\w2
\delta \psi_\mu &=& D_\mu(\omega)\epsilon -
   \ft12 m \c_\mu \epsilon\,,\label{susy}
\eea
where $D_\mu\epsilon= \partial_\mu\epsilon + \ft14 \omega_\mu^{ab}
\c_{ab}\epsilon$.  Note that the $\mu$ dependent part is separately invariant
under \eq{susy}.  In fact, this part has a larger symmetry, by being
invariant under local superconformal symmetry.
The field equations following from the Lagrangian \eq{1} are
\bea
&& \cG_{\mu\nu} + \mu^{-1} \, C_{\mu\nu} = 0\,,\label{fe1}\w2
&& R^\mu +\ft12 m \c^{\mu\nu}\psi_\nu + \ft12 \mu^{-1}\, C^\mu =0\,,\label{fe2}
\eea
up to fermionic terms in Einstein's equation, and cubic and higher than
first order in fermions in the graviton field equation, and we have used
the definitions
\bea
 \cG_{\mu\nu} &=& G_{\mu\nu} - m^2 g_{\mu\nu}\,,\label{cGdef}\w2
 G_{\mu\nu} &=& R_{\mu\nu}-\ft12 g_{\mu\nu} R\,, \label{Gdef}\w2
C_{\mu\nu} &=& \varepsilon_\mu{}^{\rho\sigma}
\nabla_\rho (R_{\sigma\nu} -\ft14 g_{\sigma\nu} R)\,,\label{Cotdef}\w2
C^\mu & = & \c^\rho \c^{\mu\nu} D_\nu R_\rho - \varepsilon^{\mu\nu\rho} (R_{\rho\sigma} -\ft14 g_{\rho\sigma} R) \c^\sigma \psi_\nu\,,
\label{CCdef}
\eea
where $C_{\mu\nu}$, which is symmetric and traceless, is the Cotton tensor
\cite{Cotton},
and the vector-spinor $C^\mu$ is supersymmetric partner, the
``Cottino vector-spinor.''  In the formulae (\ref{cGdef})--(\ref{CCdef}),
all covariant
derivatives and curvatures are defined with respect to the torsion-free
connections
$\Gamma^\mu{}_{\nu\rho}(g)$ and $\omega^{ab}(e)$, and this will be
understood in all subsequent formulae.
Next, recalling that in three dimensions
the Riemann tensor obeys the identity
\be
R_{\mu\nu}{}^{ab}= 4 e^{[a}{}_{[\mu}\,
  R_{\nu]}{}^{b]} - e^a{}_{[\mu}\,  e^b{}_{\nu]}\, R\,,
\label{riem}
\ee
where $R_\mu{}^a= R_{\mu\nu}{}^{ab} e^\nu_b$, it readily follows that
$C_\mu$ is $\c$-traceless, in the sense that $\c^\mu C_\mu=0$.

\section{The General Supersymmetric Solution}

   Here, we give a construction of the most general supersymmetric
solution to the equation \eq{fe1}.  For convenience, from this point
onwards we shall set the cosmological equal to $-1$, by taking
\be
m=1\,.
\ee
From \eq{susy}, supersymmetry implies the existence of a
solution $\ep$ to the Killing spinor equation
\be
D_\mu\ep - \ft12  \gamma_\mu \ep=0\,.\label{kseqn}
\ee
It can then be seen that the vector
\be
K^\mu = \bar\ep \gamma^\mu \ep\label{KV}
\ee
is a null Killing vector:
\be
K^\mu K_\mu=0\,,\qquad \nabla_\mu K_\nu +\nabla_\nu K_\mu=0\,.
\ee
(The null property follows from the identity $\gamma_{\mu(\alpha\beta}\,
\gamma^\mu{}_{\gamma)\delta}=0$, where $\alpha$, $\beta$, $\gamma$ and
$\delta$ are spinor indices.)
Multiplying \eq{kseqn} by $\bar\ep\gamma_\nu$ leads to the equation
\be
\nabla_\mu K_\nu  = - \ep_{\mu\nu\rho}\, K^\rho\,.\label{DKeqn}
\ee

   Using \eq{riem}, it can be seen that the integrability condition
for (\ref{kseqn}) implies
\be
\cG_{\mu\nu}\, \gamma^\nu\ep=0\,.\label{int}
\ee
Multiplying by $\bar\ep$, we see that
\be
\cG_{\mu\nu}\, K^\nu=0\,,\label{int1}
\ee
whilst multiplying instead with $\bar\ep\gamma^\rho$ gives
\be
 \ep^{\mu\nu\rho}\, K_\nu\, \cG_{\rho\sigma}=0\,.\label{int2}
\ee
The two equations \eq{int1} and \eq{int2} exhaust all the content of the
integrability condition \eq{int}.

   The existence of the null Killing vector $K^\mu$ implies that we can
choose an adapted coordinate system in which $K=\del/\del v$ and
the metric takes the form
\be
ds^2= h_{ij}\, dx^i dx^j + 2 A_i \, dx^i \, dv\,,\label{metric1}
\ee
where $h_{ij}$ and $A_i$ are independent of $v=x^0$.
   From \eq{DKeqn}, we can deduce that
\be
\ep^{\mu\nu\rho}\, \del_\nu K_\rho = 2 K^\mu\,.\label{dK}
\ee
From (\ref{metric1}), the inverse metric is given by
\be
g^{00} = - |A|^{-2}\,, \qquad g^{0i}= A^i\, |A|^{-2}\,,\qquad
g^{ij}= h^{ij} - A^i A^j |A|^{-2}\,,
\ee
where $A^i\equiv h^{ij} A_j$ and $|A|^2=A^i A_i$.  We also have
$\sqrt{-g} =\sqrt{h}\, |A|$.  Hence taking $\mu=0$ in \eq{dK},
and noting that $K_0=0$, $K_i=A_i$, we find
\be
\varepsilon^{ij} \del_i A_j = -2 \sqrt{h}\, |A|\,.\label{Aeqn}
\ee
(Our conventions are that $\ep^{\mu\nu\rho}=(1/\sqrt{-g})\,
 \varepsilon^{\mu\nu\rho}$, where $\varepsilon^{012}=-1$, and that
$\varepsilon^{12}=+1$.)

   We can choose the
$x^i$ coordinates so that $A_i dx^i = a(x^j) du$, where $u=x^1$, and then
we may take $x=x^2=a$ to be the remaining coordinate, so that
(\ref{metric1}) becomes
\be
ds^2 = h_{ij} \, dx^i dx^j + 2 x \,du \, dv\,.\label{metric2}
\ee
By means of a coordinate
transformation $v\longrightarrow v + \xi(u,x)$, the off-diagonal component
$h_{12}$ of $h_{ij}$ can be removed.  Equation \eq{Aeqn} then implies
\be
h_{22}= \fft1{4x^2}\,.
\ee
Defining $x= e^{2\rho}$, $-\infty \le \rho\le\infty$,
we therefore find that the metric (\ref{metric1})
can be cast in the form
\be
ds^2 = d\rho^2 +  2 e^{2\rho}\, du \, dv + h(u,\rho)\, du^2\,,\label{metric3}
\ee
where $h(u,\rho)$ is an undetermined function.

   Finally, we turn to the field equation.  Substituting
(\ref{metric3}) into (\ref{cGdef})--(\ref{Cotdef}), we find that the
only non-vanishing coordinate components of $\cG_{\mu\nu}$ and
$C_{\mu\nu}$ are given by
\be
\cG_{11}= -\ft12 h'' +  h'\,,\qquad
C_{11}= \ft12 h''' - \ft32  h'' + h'\,,
\ee
where a prime denotes a derivative with respect to $\rho$.  The field
equation \eq{fe1} is therefore easily solved, to give
\be
h(u,\rho) = e^{(1-\mu)\rho}\, f_1(u)  + e^{2\rho}\, f_2(u) + f_3(u)\,,
\label{hsol}
\ee
where the functions $f_1(u)$, $f_2(u)$ and $f_3(u)$ are arbitrary. It
is also manifest that the equations (\ref{int1}) and (\ref{int2})
following from the integrability condition for the Killing spinor are
satisfied.  It follows that the metric (\ref{metric3}), with
 $h(u,\rho)$ given by (\ref{hsol}), is the most general supersymmetric
solution of the theory.

There is in fact redundancy in the solution as we have presented it
above. To show this, we transform to tilded coordinates by setting
\be
\rho= \td\rho -\ft12\log a' \,,\qquad
u=a(\td u)\,,\qquad v=\td v -\ft14 e^{-2\td\rho}\, \fft{a''}{a'} +b(\td u)
\ , \label{ctrans}
\ee
where $a$ and $b$ are functions of $\td u$ and a prime denotes a
derivative with respect to $\td u$.   Choosing
$a(\td u)$ and $b(\td u)$ so that
\be
\Big(\fft{a''}{a'}\Big)' -\ft12 \Big(\fft{a''}{a'}\Big)^2
 -2 {a'}^2\, \td f_3(\td u)=0\,,\qquad b' + \ft12 a'\, \td f_2(\td u)=0\ ,
\ee
where $\td f_2(\td u)\equiv f_2(a(\td u))$ and $\td f_3(\td u)\equiv
f_3(a(\td u))$, then the functions $\td f_2$ and $\td f_3$ may be set to zero.
Furthermore, we get $\td f_1(\td u)= f_1(a(\td u))\, [a'(\td u)]^{\frac12(3+\mu)}$.  Thus, without loss of generality, we may set $f_2$ and $f_3$ to
zero in (\ref{hsol}).

   There are two special cases that must be handled carefully, when
either $\mu=1$ or $\mu=-1$.  It is easiest, though not essential,
to examine these cases before eliminating $f_2$ and $f_3$ via
the coordinate transformations (\ref{ctrans}).  It is evident from
(\ref{hsol}) that if $\mu$ approaches $+1$ then via a rescaling limit
$f_1= \td f_1/(1-\mu)$, $f_3=\td f_3 - \td f_1/(1-\mu)$, we will
find $h(u,\rho)= \rho\, \td f_1(u) + e^{2\rho}\, f_2(u) + \td f_3(u)$.

   Similarly, if $\mu$ approaches $-1$ we can take a rescaling limit
with $f_1=-\td f_1/(1+\mu)$ and $f_2= \td f_2 + \td f_1/(1+\mu)$,
to find $h(u,\rho) = \rho e^{2\rho}\, \td f_1(u) + e^{2\rho}\, \td f_2(u)
+ f_3(u)$.

   One could, of course, equivalently just solve the field equation
directly in the two special cases $\mu=\pm 1$, thereby arriving at the
same conclusions.

   In each of the cases $\mu=\pm 1$ we can then again use the coordinate
transformations (\ref{ctrans}) to remove the new $f_2$ and $f_3$ functions.
In summary, therefore, we have the most general local forms for
supersymmetric solutions
in the generic $\mu^2\ne 1$ case, and the two special cases $\mu=\pm 1$,
as follows:
\bea
\mu^2\ne1 : && ds^2 = d\rho^2 + 2e^{2\rho}\, du\, dv +
                        e^{(1-\mu)\rho}\,f(u)\,  du^2\,,\label{susysol1}\w2
\mu=1: && ds^2 = d\rho^2 + 2e^{2\rho}\, du\, dv +
           \rho\, f(u) \, du^2\,,   \label{susysol2}\w2
\mu=-1: && ds^2 = d\rho^2 + 2e^{2\rho}\, du\, dv +
         \rho\, e^{2\rho}\, f(u) \, du^2\,.\label{susysol3}
\eea
In all cases, therefore, the general supersymmetric solution is
characterised by a single arbitrary function $f(u)$.  In terms of the
coordinate $z=e^\rho$ used in the Graham-Fefferman discussion of
boundary conditions, the $\mu=\pm1$ solutions have logarithmic
singularities.

    A supersymmetric
solution of the form \eq{metric3} and \eq{hsol}, but with
$f_1$, $f_2$ and $f_3$ restricted to be constants,
was found in \cite{dersar}.  It was later observed in \cite{olsate}
that these constants could be replaced by arbitrary functions of the
variable we are calling $u$, but it was conjectured (erroneously) that
the solutions would then cease to be supersymmetric.  The fact that
$f_2$ and $f_3$ can always be removed by coordinate transformations was
not noted in \cite{dersar,olsate}.  Fully non-linear chiral pp-waves have
also been obtained in \cite{comgrav6}.  They discuss three cases, which
correspond to taking just one of our functions $f_2$, $f_3$ or
$f_1$ to be non-zero respectively.  The first two cases are also solutions
of pure Einstein gravity and hence may be reduced to a locally-AdS$_3$
solution by a gauge transformation (as we showed above). The third
solution coincides with our \eq{susysol1} if $\mu^2\ne 1$.  If $\mu=\pm 1$,
their solution corresponds to our \eq{susysol2} or \eq{susysol3}.

   There has been some controversy about the degrees of freedom of
topologically massive gravity.  Clearly, these supersymmetric solutions
exhibit the same number of local degrees of freedom for all values of
the Chern-Simons coupling constant $\mu$, including the chiral values
$\mu=\pm 1$.  We shall discuss later whether or not these local
solutions are globally well-defined.  We shall see shortly that the
supersymmetric solutions we have obtained exhaust the class of all
solutions admitting a null Killing vector field.  This latter condition
is often taken to be the criterion for having a ``gravitational wave.''
This again points strongly to the conclusion that the local degrees
of freedom are the same for all values of the coupling constant $\mu$.

   In all three cases, the Cotton tensor $C_{\mu\nu}$ can be written simply
in terms of the Killing vector $K$:
\bea
\mu^2\ne 1:&& C_{\mu\nu} = \ft12 \mu (1-\mu^2) f(u)\, e^{-(3+\mu)\rho}\,
        K_\mu\, K_\nu\,,\nn\\
\mu=+1:&&  C_{\mu\nu} = f(u)\, e^{-4\rho}\, K_\mu\, K_\nu\,,
\label{cottons}\\
\mu=-1:&& C_{\mu\nu}= f(u) \, e^{-2\rho}\, K_\mu\, K_\nu\,.\nn
\eea
Since the vanishing of the Cotton tensor is the criterion for a
three-dimensional metric to be conformally flat (and hence locally
equivalent to AdS$_3$), we see that the metrics we have constructed
are inequivalent to AdS$_3$ whenever $f(u)$ is non-zero.

\subsection{The Killing spinor}

   In order to construct the Killing spinor explicitly in the supersymmetric
solutions obtained above, it is useful to introduce an orthonormal frame
for the metric \eq{metric3}.  We shall take
\be
e^0= e^{2\rho-\beta}\, dv\,,\qquad e^1= e^\beta\, du + e^{2\rho-\beta}\, dv\,,
\qquad e^2=d\rho\,,
\ee
where we have written $h(u,\rho)=e^{2\beta}$.  The torsion-free spin
connection is then given by
\bea
\omega_{01}&=& \dot\beta e^{-\beta}\, (e^0-e^1)-(\beta'-1)\, e^2=
    -\dot\beta\, du -(\beta'-1)\, d\rho\,,\nn\\
\omega_{02}&=& (\beta'-1)\, (e^0-e^1) -  e^0 = -(\beta'-1)\, e^\beta\, du
               -  e^{2\rho-\beta}\, dv\,,\label{spincon}\\
\omega_{12} &=& -\beta'\, (e^0-e^1) +  e^0 = \beta'\, e^\beta\, du +
                e^{2\rho-\phi}\, dv\,,
\eea
where, as usual, $\dot\beta =\del\beta/\del u$ and $\beta'=\del\beta/\del \rho$.
The Lorentz-covariant exterior derivative on spinors is given by
\be
D= d+\ft14 \omega_{ab}\, \gamma^{ab}\,.
\ee

   A convenient, real, basis for the Dirac matrices is provided by taking
\be
\gamma_0= \im \sigma_2\,,\qquad \gamma_1= \sigma_1\,,\qquad
   \gamma_2= \sigma_3\,,
\ee
where $\sigma_i$ are the standard Pauli matrices.  From \eq{spincon} we
therefore find that the Killing spinor equation \eq{kseqn}, \ie
$D\ep -\ft12 \gamma_a\ep\, e^a=0$, becomes
\be
d\ep + e^{2\rho-\beta}\, (\sigma_1+\im \sigma_2)\ep\, dv +
  \ft12 [\dot\beta\, \sigma_3 -\beta'\, (\sigma_1+\im\sigma_2)]\ep\, du
   +\ft12(\beta'-2)\,\sigma_3\ep\, d\rho=0\,.
\ee
This is solved by a $v$-independent spinor satisfying
$(\sigma_1+\im\sigma_2)\ep=0$, with
\be
\ep=\begin{pmatrix} \alpha \\
                    0\end{pmatrix}\,,\qquad \alpha= h^{-1/4}\, e^{\rho}\,.
\ee

    Note that if the function $f(u)$ is taken to be zero, the metric
reduces to
\be
ds^2 = d\rho^2 + 2 e^{2\rho}\, du\, dv= d\rho^2 + e^{2\rho}\, (-dt^2+dx^2)\,,
\label{poincare}
\ee
which is AdS$_3$ in the Poincar\'e patch.  This has a supersymmetry
enhancement, with two Killing spinors given, in the basis
$e^0=e^{\rho} dt$, $e^1=e^{\rho} dx$, $e^2=d\rho$,  by
\be
\ep_1 = e^{\ft12 \rho} \, \begin{pmatrix} 1 \\ 0\end{pmatrix}\,,\qquad
\hbox{and} \qquad \ep_2= \Big( e^{-\ft12 \rho} +
   e^{\ft12 \rho}\, (\im t \sigma_2 + x \sigma_1)\Big)
   \begin{pmatrix} 0 \\ 1\end{pmatrix}\,.\label{pks}
\ee

\subsection{Local degrees of freedom and
general solution with null Killing vector}

   As we indicated above, the existence of a null Killing vector field is
often taken to be the defining property of a gravitational wave.
It is of interest
therefore to examine the more general class of solutions
that we can obtain if we only make the assumption that there exists a
null Killing vector, without the further property \eq{DKeqn} that followed
from supersymmetry.  We may still, without loss of generality, choose
coordinates so that the metric takes the form \eq{metric2}, and
make a further coordinate transformation of the form $v\longrightarrow
v+\xi(u,x)$ to remove the off-diagonal component $h_{12}$.  Thus we may
take the metric to be of the form
\be
ds^2= h_1(u,x) du^2 + h_2(u,x) dx^2 + 2 x \, du dv\,.
\ee

   Substituting into the field equation \eq{fe1}, we first find that
\be
h_2 = \fft{1}{4 x^2}\,.
\ee
It is convenient, as in the supersymmetric case, to introduce a new
coordinate $\rho$, related to $x$ by $x=e^{2\rho}$.  We then find that
for generic values of $\mu\ne \pm 1$, the solution is given by
\be
ds^2= d\rho^2 + 2 e^{2\rho}\, du \, dv + [e^{(1 -\mu)\rho}\, f_1(u)
   + e^{2\rho}\, f_2(u) + f_3(u)]\, du^2\,,
\ee
where the functions $f_1(u)$, $f_2(u)$ and $f_3(u)$ are arbitrary.
As in the supersymmetric case we discussed previously, the functions
$f_2(u)$ and $f_3(u)$ can be eliminated by means of further coordinate
transformations.  Also, as for the supersymmetric solutions, special
cases arise at $\mu=\pm 1$.  After eliminating the redundant functions in
all the cases, we arrive at the general solutions:
\bea
\mu^2\ne 1: && ds^2 = d\rho^2 + 2e^{2\rho}\, du\, dv +
                        e^{(1- \mu)\rho}\,f(u)\,  du^2\,,\label{nullsol1}\w2
\mu=1: && ds^2 = d\rho^2 + 2e^{2\rho}\, du\, dv +
           \rho\, f(u) \, du^2\,, \label{nullsol2}\w2
\mu=-1:  && ds^2 = d\rho^2 + 2e^{2\rho}\, du\, dv +
         \rho\, e^{2\rho}\, f(u) \, du^2\,.\label{nullsol3}
\eea
These are precisely the same as the supersymmetric solutions we found
in (3.20), (3.21) and (3.22.  Note, however,
that we can also obtain non-supersymmetric solutions with a null Killing
vector, by reversing the orientation, specified by the sign of
$\varepsilon^{012}$. This amounts to reversing the sign of $\mu$ in the
solutions (3.36), (3.37) and (3.38). Thus the solutions having just a null Killing vector
are either supersymmetric as they stand, or they would be supersymmetric if
the opposite sign choice for the parameter $\mu$ in theory
were taken.

   It is interesting to note that all the solutions we have constructed,
both the supersymmetric ones and the more general solutions
\eq{nullsol1}--\eq{nullsol3}, have the feature that they are written in
Kerr-Schild form.  To see this, we observe that the null Killing vector
$K=\del/\del v$, if written as a 1-form $K= K_\mu dx^\mu$, is simply
given by $K= e^{2\rho}\, du$.  Thus the solutions can all be written in
the form
\be
g_{\mu\nu} = g_{\mu\nu}({\rm AdS}) + s(u,\rho)\, K_\mu \, K_\nu\,,
\label{kerrschild}
\ee
where the function $s(u,\rho)$ is given by
\be
s(u,\rho) = h(u,\rho) \, e^{-4\rho}\,.
\ee
The metric $ g_{\mu\nu}({\rm AdS}) $ is just the AdS$_3$ metric on the
Poincar\'e patch, written in the form
\be
ds^2({\rm AdS}) = d\rho^2 + 2 e^{2\rho}\, du\, dv\,.
\ee

\subsection{Kerr-Schild form and vanishing quantum corrections}

Since a null Killing vector is necessarily geodesic, satisfying
$K^\nu\nabla_\nu K_\mu=0$, it follows that all the conditions for
\eq{kerrschild} to be a Kerr-Schild metric are fulfilled.  In particular,
this means that if one considers a ``linearised approximation'' in which
the function $h(u,\rho)$ is taken to be small (recall that it has a
factor $f(u)$, where $f$ can be chosen arbitrarily), then the linear
approximation is in fact exact.

    An important  consequence of the Kerr-Schild form of the metric,
 encapsulated
in equation \eq{cottons}, is that $C_{\mu\nu}$ and the metric $g_{\mu\nu}$
are the only conserved symmetric tensors that can be constructed from
polynomials in the metric and the Ricci tensor and its covariant derivatives.

To see this, note that since we are in three dimensions,
when constructing symmetric conserved tensors
we need only consider the
Ricci tensor and its covariant derivatives. But by the field equations
this is proportional to the Cotton tensor and so we may instead
consider only the
Cotton  tensor and its covariant derivatives. However,
whenever one takes a covariant derivative
of the Cotton tensor, or of one of its covariant derivatives to
arbitrary order,
on adds a further multiple of the null Killing
vector $K_\mu$. Thus the contractions necessary
to construct a second rank symmetric tensor from
arbitrary powers of  covariant derivatives of arbitrary
order of the Cotton tensor
 must necessarily contain at least one, and in general many,
factors of the vanishing quantity $K_\mu K^\mu$.

In the language of \cite{cogihepo}, the supersymmetric solutions are
therefore {\it universal}.  As explained in \cite{cogihepo}, if one envisages
quantum corrections to the classical
equations of motion \eq{fe1}, in which the right-hand side is replaced by
a symmetric conserved tensor constructed from a finite number of polynomial
terms, each of which is a monomial in the Riemann  tensor and its
covariant derivatives, then these supersymmetric solutions will still
solve the corrected equations, possibly with a shifted value for the
Chern-Simons coupling constant $\mu$.

These considerations do not, of course, preclude the construction
of non-zero non-polynomial invariants, as recently discussed in \cite{Page}.

\section{The Solutions in Global Coordinates}

   The solutions we have presented are written in terms of coordinates
on the Poincar\'e patch of AdS$_3$.  In order to study the behaviour
on the boundary and on the horizon, it is useful to work in a global
coordinate system for AdS$_3$.  To do this, we introduce embedding
coordinates $X^A$, for $A=0,1,2,3$, on the hyperboloid
\be
\eta_{AB}\, X^A X^B= -1  \,,\qquad \eta_{AB}=\hbox{diag}(-1,1,1,-1)\,.
\ee
We then introduce the global coordinates $(\tau,r,\phi)$
\bea
X^0&=&  \sqrt{1 +r^2}\, \cos\tau\,,\qquad
X^3 = \sqrt{1+r^2}\, \sin\tau\,,\nn\w2
X^1&=&  r\, \cos\phi\,,\qquad \qquad
X^2=    r\, \sin\phi\,.\label{Xglob}
\eea
The AdS$_3$ metric is given by
\be
ds^2({\rm AdS}) = \eta_{AB} dX^A dX^B = -(1+r^2) d\tau^2
   + \fft{dr^2}{1+r^2} + r^2 d\phi^2\,.
\ee

   The relation between the coordinates $(v,u,\rho)$ of the Poincar\'e
patch and the embedding coordinates $X^A$ is
\bea
X^0-X^1 &=&  e^{\rho}\,,\qquad X^0+X^1=
e^{-\rho} +2 u v \, e^{\rho}\,,\nn\w2
X^2 &=& \fft{v+u}{\sqrt2}\, e^{\rho}\,,\qquad
X^3= \fft{v-u}{\sqrt2}\, e^{\rho}\,.\label{Xpoin}
\eea
Comparing (\ref{Xglob}) and (\ref{Xpoin}), we obtain the coordinate
transformation
\bea
v&=& \fft{r\sin\phi +
   \sqrt{1 +r^2}\, \sin\tau}{\sqrt2\,(\sqrt{1 +r^2}\, \cos\tau
         -r\cos\phi)}\,,\qquad
u=\fft{r\sin\phi - \sqrt{1+r^2}\, \sin\tau}{
        \sqrt2\, (\sqrt{1+r^2}\, \cos\tau
         -r\cos\phi)}\,,\nn\w3
e^{\rho} &=& \sqrt{1+r^2}\, \cos\tau  - r\cos\phi\,,\label{poinglob}
\eea
which relates the global and Poincar\'e coordinates.

   We now investigate whether the supersymmetric solutions admits a smooth
conformal compactification.  Consider the metric (\ref{susysol1}).  This
can be written as
\be
ds^2 = r^2\, \Big( - (1+r^{-2}) d\tau^2 + \fft{dr^2}{r^2 (1+r^2)} +
   d\phi^2 + \fft{1}{r^2}\, e^{(1-\mu)\rho}\, f(u) du^2\Big)\,.
\ee
Let $r=1/y$.  Then $ds^2= \Omega^{-2}\, \td g_{\mu\nu} dx^\mu dx^\nu$, with
$x^\mu=(y,\tau,\phi)$ and $\Omega= y$.  The conformal boundary is at $y=0$,
and $d\Omega\ne0$.
If $\td g_{\mu\nu}$ has a smooth extension across the surface $y=0$, then
the metric will have a smooth conformal compactification
\cite{hawkelli,wald}.   If $1+\mu\ge0$,
this will be true for all values of $\tau$ and $\phi$ except those for
which $\cos\tau=\cos\phi$.  Thus, the metric certainly has a smooth
conformal compactification in a single Poincar\'e patch.  However, the
metric $\td g_{\mu\nu}$ does not have a smooth extension across the points
for which $\cos\tau=\cos\phi$.  From the strong directional dependence of
the metric functions $\td g_{\mu\nu}$, it is extremely plausible
that no other choice of asymptotic coordinates would yield a smooth
compactification across the points where $\cos\tau=\cos\phi$.

   We shall
not attempt a formal, rigorous, mathematical proof here, but merely note
that from \eq{poinglob},
\bea
u &=& \fft{\cos p}{\sqrt2\,\sin p} +
              \fft{\cos q}{4\sqrt2\,  r^2\sin q\, \sin^2 p} +
                      {\cal O}\Big(\fft1{r^4}\Big)\,,\nn\w2
e^{\rho} &=& -2 r \sin p\, \sin q + \fft{\cos(p+q)}{2 r} +
            {\cal O}\Big(\fft1{r^3}\Big)\,,\nn\w2
du^2 &=& \fft{dp^2}{2 \sin^4 p} + \fft{ dp(\sin p\, dq +
\cos p\, \sin2q\, dp)}{ 4 r^2
                  \sin^5p \, \sin^2 q} + {\cal O}\Big(\fft1{r^3}\Big)\,,
\eea
where we have defined
\be
\tau= p+q\,,\qquad \phi=p-q\,.
\ee
It seems clear that, no matter what choice we make for $f(u)$, the metric
will contain singularities when $\sin p$ or $\sin q$ vanishes.

\section{Nester-Witten Energy}

In this section we shall obtain a Nester-Witten identity for
topologically massive gravity which
will allow us to identify the energy and angular momentum which appear
in it   and to attempt to  establish a positivity property.
Because the supersymmetry variations
are the same as for the theory without the Chern-Simons term,
one might anticipate similar difficulties to those encountered by
Boulware, Deser and Stelle \cite{Boulware1,Boulware2}
in higher-derivative four-dimensional
gravity. This is indeed what we find.  It is easy to establish the
relevant Nester-Witten identity, but since the bulk term contains the
Cotton tensor, whose sign is indefinite, we are unable to establish a
general positive-energy property for solutions of topologically massive
supergravity.

   From the supercurrent $J^\mu(\ep_1) =\nabla_\nu(\bar\ep_1
 \, \gamma^{\mu\nu\rho}\, \psi_\rho)$, we define, via its variation
\be
\delta_{\ep_2}\, J^\mu(\ep_1) = \nabla_\nu(\bar\ep_1 \, \gamma^{\mu\nu\rho}\,
\delta_{\ep_2}\psi_\rho)= \nabla_\nu(\bar\ep_1\,
   \gamma^{\mu\nu\rho}\, \hat\nabla_\rho \ep_2)\,,
\ee
the quantity
\be
X\equiv \int_\Sigma \delta_{\ep_2}\, J^\mu(\ep_1) d\Sigma_\mu=
 \int_\Sigma  \nabla_\nu(\bar\ep_1\,
   \gamma^{\mu\nu\rho}\, \hat\nabla_\rho \ep_2)d\Sigma_\mu\,.\label{LHS}
\ee
Using Stokes' theorem, this may be re-expressed as
\be
X= \ft12 \oint_{\del\Sigma} (\bar\ep_1\,
   \gamma^{\mu\nu\rho}\, \hat\nabla_\rho \ep_2)d\Sigma_{\mu\nu}\,.
\label{RHS}
\ee

   After some algebra, in which we make use of the facts that
\bea
\hat\nabla_\mu \ep &=& \nabla_\mu -\ft12  \gamma_\mu \ep\,,\qquad
\hat\nabla_\mu\bar\ep= \nabla_\mu\bar\ep +\ft12  \bar\ep\gamma_\mu\,,\nn\\
\gamma^{\mu\nu} &=& \ep^{\mu\nu\rho}\, \gamma_\rho\,,\qquad
\gamma^{\mu\nu\rho}=\ep^{\mu\nu\rho}\,,
\eea
and \eq{riem},
we find that the expression (\ref{LHS}) for $X$ may be written as
\be
X= \int_\Sigma\Big(\hat\nabla_\nu \bar\ep_1\,
  \gamma^{\mu\nu\rho}\, \hat\nabla_\rho \ep_2  +
  \ft12 \cG_\nu{}^\mu \bar\ep_1 \gamma^\nu\ep_2\Big) d\Sigma_\mu\,,
\label{LHS2}
\ee
where $\cG_{\mu\nu}$ is the Einstein tensor with cosmological term,
as defined in (\ref{Gdef}) and (\ref{cGdef}).

   One may argue that the energy $E[\xi]$, where $\xi$ is a Killing vector
with spinorial square root:
\be
\xi^\nu =\bar\ep\gamma^\nu\ep\,,
\ee
is given by
\be
E[\xi]= -\fft1{4\pi G}\, X  =     \fft1{8\pi G}\, \oint_{\del\Sigma}
 \bar\ep \gamma^{\mu\nu\rho}\, \hat\nabla_\rho
\ep\, d\Sigma_{\mu\nu}\,,\label{EX}
\ee
where now we take $\ep_1=\ep_2=\ep$.
To see this, we consider a ``deformed''
metric that satisfies the Einstein equations
\be
\cG_{\mu\nu} = 8\pi G\, T_{\mu\nu}\,,
\ee
and that is close to a ``vacuum'' background metric
satisfying $\cG_{\mu\nu}=0$.  The vacuum is assumed to admit a Killing
spinor $\ep$, so $\hat\nabla_\mu\ep=0$ in the background.  Taking
$\ep_1=\ep_2=\ep$ in (\ref{LHS2}), evaluated in the deformed metric,
we see that the first term is quadratically small since $\hat\nabla\ep$
itself is linearly small.  Thus for a sufficiently small deformation,
$X$ in (\ref{LHS2}) can be taken to be given by
\be
X= 4\pi G \int_\Sigma T_{\mu\nu} K^\nu\, d\Sigma^\mu\,.
\label{LHS3}
\ee

    The standard way of proceeding is now to impose on the spinor field
$\ep$ the Witten condition $\gamma^i\hat\nabla_i\ep=0$, where the index
$i$ indicates quantities in the surface $\Sigma$.  Subject to the Witten
condition, the first term in $X$ in \eq{LHS2} is negative semi-definite,
as desired.  However, the second term in \eq{LHS2} is more problematic.
From \eq{EX} and the field equation \eq{fe1}, we have
\be
E[K] \ge \fft1{8\pi\mu G}\, \int_\Sigma C_\nu{}^\mu \xi^\nu\,
   d\Sigma_\mu\,.
\ee
However, because $C_\nu{}^\mu$ is third order in derivatives of the
metric, it will in general have no definite sign.  This is clear from our
plane-wave solution, whose Cotton tensor is given by (\ref{cottons}).  We
can take the arbitrary function $f(u)$ to have either sign.  We reluctantly
conclude that our Witten identity is incapable of providing an answer
to the question of whether excitations around an AdS$_3$ background must
always have positive energy.

\subsection{The BTZ black hole, and its Killing spinors}

    Although there are difficulties in applying the Nester-Witten
identity to a general solution of topologically massive gravity, it can
be applied successfully to the BTZ black hole, for which the Cotton tensor
vanishes.  BTZ black holes are locally AdS$_3$, but with identifications
which imply that, depending on the choice of the mass and
angular momentum parameters, some or all of the supersymmetry is
broken.  The metric takes the form
\be
ds^2= - U(r) dt^2 + \fft{dr^2}{U(r)} +
   r^2\Big(d\phi -\fft{J}{2r^2}\, dt\Big)^2\,,\qquad
 U(r) = r^2 -M + \fft{J^2}{4 r^2}\,.
\ee
When $M=J=0$, this has the same form as (\ref{poincare}), with $r=e^\rho$
and $x=\phi$.  Since $\phi$ is identified modulo $2\pi$, we see from
(\ref{pks}) that although the Killing spinor $\ep_1$ survives, $\ep_2$
does not. Thus the vacuum limit of the BTZ black hole has one half of
maximal supersymmetry, with just one Killing spinor, given by
\be
\ep= \fft{r^{1/2}}{\sqrt2} \begin{pmatrix} 1\\ 1\end{pmatrix}\,.
\label{btzks}
\ee

  When $M$ and $J$ are non-zero, a straightforward calculation shows that
the Killing spinor exterior derivative $\hat D= D-\ft12 \gamma_a e^a$ is
given by
\be
\hat D= d + \ft12 \Big[\Big(r+ \fft{J}{2 r}\Big) \sigma_3 -
 U^{1/2}\, \im\sigma_2\Big] (dt-d\phi) + \ft12 U^{-1/2}\,
    \Big(\fft{J}{2r^2}-1\Big) \sigma_1\, dr\,.\label{hatD}
\ee
 From this, we can solve locally for the Killing spinors satisfying
$\hat D\ep=0$, finding
\be
\ep =e^{-g(r) \sigma_1}\, e^{(\phi-t) P}\, \ep_0\,, \label{ksbtz}
\ee
where $\ep_0$ is an arbitrary constant spinor and
\be
   P\equiv \ft14(M+J+1)\sigma_3 +\ft14(M+J-1)\im\sigma_2\,.
\ee
The function $g(r)$ is given by
\be
e^{4g(r)} = \fft{(e^{2\rho} + M-J)}{e^{2\rho}(e^{2\rho} +M+J)}\,,
\ee
where we have introduced the new radial variable
 $\rho$ such that $U^{-1/2}\, dr =d\rho$:
\be
r^2 = \ft14 [(e^{2\rho}+M)^2 -J^2]\, e^{-2\rho}\,.
\ee
Although (\ref{ksbtz}) gives the expected two local solutions of the
Killing spinor equation for arbitrary $M$ and $J$ (as it must since
locally the BTZ black hole is just AdS$_3$), we see that the global
requirement that $\phi$ be periodic (with period $2\pi$) in the BTZ
solution eliminates both of the Killing spinors in general.  Since
$\det P=-\ft14(M+J)$, we see that an
exception arises in the extremal case
\be
J=-M\,,\label{extremal}
\ee
for which (\ref{ksbtz}) gives the single globally-defined Killing spinor
\be
\ep = e^{-g(r)}\, \ep_0\,,\qquad
\hbox{with}\qquad \ep_0=\fft1{\sqrt2}\,\begin{pmatrix} 1\\ 1\end{pmatrix}\,.
\ee

\subsection{Nester-Witten charge of the BTZ black hole}

   The fact that the BTZ vacuum admits the single Killing spinor
(\ref{btzks})
implies that there is just one Killing vector field that has a spinorial
square root,\footnote{This should be contrasted with the situation for
AdS black holes in dimensions $D>3$, where any Killing vector of the vacuum
can be expressed as a linear combination of Killing vectors with spinorial
square roots.  This difference stems from the fact that in dimensions
$D>3$ the black holes are asymptotic to AdS$_D$ globally, whereas in $D=3$
the BTZ black hole is asymptotic to AdS$_3$ only locally.}
namely
\be
K=\bar\ep \gamma^\mu \ep \del_\mu = \fft{\del}{\del t}
   -\fft{\del}{\del\phi}\,.
\label{Kbtz}
\ee

  The spinor field used in the Nester-Witten construction must tend
asymptotically to a Killing spinor of the BTZ vacuum.  Since the BTZ
vacuum has just the one Killing spinor \eq{btzks}, it follows that there
is only one choice of boundary conditions for spinor fields used in the
Nester-Witten construction, when applied to the general BTZ black hole.
Therefore, we can apply \eq{EX} only to the Killing field \eq{Kbtz}.
We find that the associated Nester-Witten charge is given by
\be
E\left[\fft{\del}{\del t} - \fft{\del}{\del \phi}\right] = M+J\,.
\ee
Note that this vanishes, as one would expect, in the extremal limit
(\ref{extremal}).

\section{Conclusion}

    In this paper, we have constructed the most general fully
non-linear solutions of
topologically massive supergravity admitting a null Killing vector
field.  These exhaust the class of all supersymmetric solutions of the
theory.  We have established a Nester-Witten identity, allowing us to
read off the Nester-Witten charges appearing in the superalgebra.
Because the Cotton tensor appears in the bulk term of the identity,
one cannot use this method to establish a general positivity property
for the Nester-Witten charges, except in case such as that of the
BTZ black hole, for which the Cotton tensor vanishes.  Our solutions
indicate that the local degrees of freedom of topologically massive
supergravity are the same for all values of the Chern-Simons
coupling constant $\mu$.  However, since our solutions are not globally
AdS, we are unable to tackle the interesting and controversial
question of their relationship to the linearised solutions
used to construct representations of the isometry
group $SO(2,2)$.  It would be interesting to investigate further
the connection between the global supersymmetry algebra, the
Nester-Witten identity, and the boundary conditions satisfied by an
appropriate set of solutions.

\section*{Acknowledgements}

The research of C.N.P. is
supported in part by DOE grant DE-FG03-95ER40917.  The research of E.S. is
supported in part by NSF grant PHY-0555575.  C.N.P. and E.S. are grateful
to Cambridge-Mitchell Collaboration for financial support, and the
Cambridge Centre for Theoretical Cosmology, DAMTP, for hospitality during
the course of this work. We would like to thank Yoshiaki Tanii and David Chow for bringing to
our attention some errors in the earlier version of this paper.


\end{document}